\documentclass[a4paper,11pt]{iopart}

\expandafter\let\csname equation*\endcsname\relax

\expandafter\let\csname endequation*\endcsname\relax
\usepackage{amsfonts}
\usepackage{amsmath,amssymb}
\usepackage{hyperref}

\usepackage{graphicx}
\usepackage{cite}
\usepackage{color}
\usepackage{srcltx}
\usepackage{subfigure}
\def \hh{\mathcal{I}}
\def \COPc{\textrm{COP}_c}
\def \COPps{\textrm{COP}_{ps}}
\def \COP{\textrm{COP}}

\begin{document}

\title{Cyclic active refrigerators}
\author{S. Liu and A. Datta and A. C. Barato}
\address{Department of Physics, University of Houston, Houston, Texas 77204, USA}
\ead{barato@uh.edu}

\begin{abstract}
Thermodynamic cycles are idealized processes that can convert heat into work or produce heat flow against a temperature gradient with the input of work. 
They remain an active area of research in modern stochastic thermodynamics.  In particular,  cyclic active heat engines have been shown 
to display a rich phenomenology, such as ``violations'' of the Carnot bound on efficiency and an improved performance in comparison to their passive counterparts. 
We introduce the concept of cyclic active refrigerators using a previously derived second law for cyclic active systems. 
We show that for cyclic active refrigerators, a naive definition of the  coefficient of performance can exceed the bound set 
by the standard second law for passive refrigerators. We also show that cyclic active systems can behave like a Maxwell's demon,
with heat flowing from the cold to the hot reservoir without any work input. Beyond this phase, cyclic active systems can enter a hybrid phase, 
functioning as both a heat engine and a refrigerator simultaneously. Our results are obtained with two models that involve active Brownian particles, 
a simpler one  that allows for analytical results and a more realistic one that is analyzed through numerical simulations.
 \end{abstract}

\section{Introduction}

Stochastic thermodynamics \cite{seif12,seif25} is a modern theory that generalizes thermodynamics to systems that can be small and operate out of equilibrium.
In marked contrast to standard thermodynamics, the entropy change (and related quantities such as heat and work) is defined 
as a fluctuating observable. Fluctuation relations \cite{evan93,gall95,jarz97,croo99,lebo99,seif05}, which arguably jump-started the field of stochastic thermodynamics, can be expressed as a 
symmetry in the probability distribution of this fluctuating entropy. In this context, the second law of thermodynamics, which is valid for the average entropy, is a corollary of fluctuation relations. 
Relevant topics of research  within stochastic thermodynamics include the relation between information and thermodynamics \cite{parr15}, 
the thermodynamic uncertainty relation (and related bounds) \cite{bara15,ging16,piet16,bara16,piet22,ohga23,meie25}, the relation between precision and dissipation in biophysics 
\cite{qian07,lan12,meht12,gove14a,zhan20,junc20a,rana20,hath24,gent25}, and the inference of entropy production from incomplete information \cite{rold10,mart19,kim20,otsu20,skin21,meer22,haru22}.

The particular topic we address here is the study of cyclic heat engines and refrigerators. Thermodynamic cycles were a main motivation for the development of 
standard thermodynamics. They remain an active topic of research within stochastic thermodynamics, with the novelty that systems can be small and operate in finite time, 
overcoming two main limitations of standard thermodynamics, which deals with macroscopic systems undergoing quasi-static protocols.
Specifically, a model of a small cyclic heat engine as a colloidal particle in a harmonic potential has been proposed in \cite{schm08} and later realized experimentally in \cite{blic12}. 
Further theoretical work\cite{espo10,izum11,tu14,bran15,raz16,ray17,koyu19} and experiments  \cite{stee11,mart15,mart16,ross16,mart17} have been done on these passive heat engines. 
We point out that finite time heat engines beyond standard thermodynamics,  have been studied within linear response theory before the emergence of stochastic thermodynamics \cite{curz75,band82,andr84,leff87}.

The concept of a cyclic active heat engine was proposed in an original experiment \cite{kris16}. Similar to the scenario from \cite{blic12}, this experiment 
was made with a colloidal particle under the action of a harmonic potential. However, the bath  was active due to the presence of several bacteria with a 
size comparable to the size of the colloidal particle. This paper contains several interesting observations, such as the work extracted by an active heat engine can 
be larger than the work extracted by its passive counterpart (without the bacteria in the bath) and the ``efficiency'' can be larger than the Carnot bound.

This experiment has led  to much work on cyclic active heat engines  \cite{zaki17,saha19,ekeh20,holu20,holu20b,kuma20,lee20,fodo21,gron21,alba23,wies24}. More generally, the relation between active systems and stochastic thermodynamics 
has received much attention \cite{argu16,piet18a,mand17,spec18,shan18,piet19,dabe19,gopa21,mark21,capr23,frit23,bebo25,dabe25}.  A key theoretical contribution was the development of the appropriate statement 
of the second law for cyclic active heat engines \cite{datt22}. This second 
law provides a bound on extracted work that does not contain the dissipation due to the bacteria and only has quantities that depend on the observable degrees of freedom, i.e., 
the position of the colloidal particle for the experiment in \cite{kris16}. An extra information theoretic term emerges in this second law, 
as compared to the second law for passive heat engines, which accounts for observations such as the ``violation'' of the Carnot bound from \cite{kris16}. 

In this study, we consider a different phase of operation of a cyclic active system, i.e., we propose the study of an active refrigerator using the statement of the 
second law derived in \cite{datt22}. We show that a naive definition of the coefficient of performance of an active refrigerator can exceed the standard 
bound for a passive refrigerator, similar to the behavior of the efficiency of an active heat engine. Beyond this phase, a cyclic active system  can operate like
Maxwell's demon, i.e., with heat flowing from cold to hot with no input of work. Furthermore, we show that an active system has a hybrid phase where it 
operates as a heat engine and a refrigerator at the same time, i.e., heat flows from the cold to the hot reservoir and work is extracted. Our results are obtained with two models. 
One with an active particle in the presence of a harmonic potential, which allows for analytical calculations. The other model has a passive particle, also in the presence of a harmonic potential, 
that interacts  with several hidden active particles. This second model is analyzed numerically and is closer to the experiment from \cite{kris16}. 

The paper is organized as follows. In Sec. \ref{sec2}, we present the general framework and the second law for cyclic active systems. 
In Sec. \ref{sec3}, we use the second law for cyclic active systems to speculate on possible phases of operation for active refrigerators. 
These phases of operation are then realized with two models. One in Sec. \ref{sec4}, which allows for analytical calculations. 
The other in Sec. \ref{sec5}, which is more realistic and is analyzed numerically. We conclude in Sec. \ref{sec6}. There are two appendices 
with details about the two models. \ref{app1} contains the calculations for the first model and \ref{app2} contains the definition of the second model.

\section{Second law for cyclic active systems}
\label{sec2}

\subsection{General Framework}

One postulate of stochastic thermodynamics is that the system dynamics is Markovian, i.e., the probability of state $i$
at time $t$ follows the master equation
\begin{equation}
\frac{d}{dt}P_i(t)= \sum_j\left(P_j(t)w_{ji}(t)-P_i(t)w_{ij}(t)\right),
\label{eqM}
\end{equation}
where $w_{ij}(t)$ is the time-dependent transition rate from $i$ to $j$. Since we consider cyclic systems, the transition rates have a period $\tau$, i.e., $w_{ij}(t)=w_{ij}(t+\tau)$. 
In the long-time limit, the probability becomes periodic with a period $\tau$. We are interested in this long-time limit behavior and we denote the long-time limit solution 
of the master equation simply by $P_i(t)$, which fulfills $P_i(t)=P_i(t+\tau)$. A quantity of interest here is the probability current at time $t$, which is defined as 
\begin{equation}
J_{ij}(t)\equiv P_i(t)w_{ij}(t)-P_j(t)w_{ji}(t).
\label{eqprobcurr}
\end{equation}
This quantity gives the average number of jumps per time from $i$ to $j$ minus the number of jumps per time from $j$ to $i$, at time $t$.

 Another postulate of stochastic thermodynamics is that the transition rates are related to thermodynamic parameters characterizing the external reservoir through the generalized detailed balance relation \cite{seif12}. These time-periodic thermodynamic 
 parameters are the inverse temperature $\beta(t)$, the free energy of state $i$ written as $E_i(t)$ and the thermodynamic affinity $F(t)$. 
Freezing the protocol at time $t$ and letting the system evolve to a stationary state, a non-zero F(t) implies evolution towards a nonequilibrium stationary state. 
If $F(t)$ is zero, this stationary state would be the equilibrium one.  
 This affinity $F(t)$ is the quantity that makes the cyclic system active, i.e., if F(t)=0 for all $t\in[0,\tau]$ then the cyclic system is passive \cite{datt22}.
 In general, there can be more than one affinity. However, we will consider models that only have one thermodynamic affinity.  The generalized detailed balance relation reads 
 \begin{equation}
 \ln\frac{w_{ij}(t)}{w_{ji}(t)}= -\beta(t)[E_j(t)-E_i(t)]+F(t)d_{ij},
 \label{eqgdb}
 \end{equation}
where the $d_{ij}$ are generalized distances that are anti-symmetric, i.e., $d_{ij}= -d_{ji}$.

\begin{figure}
\centering
\includegraphics[angle =270,width=85mm]{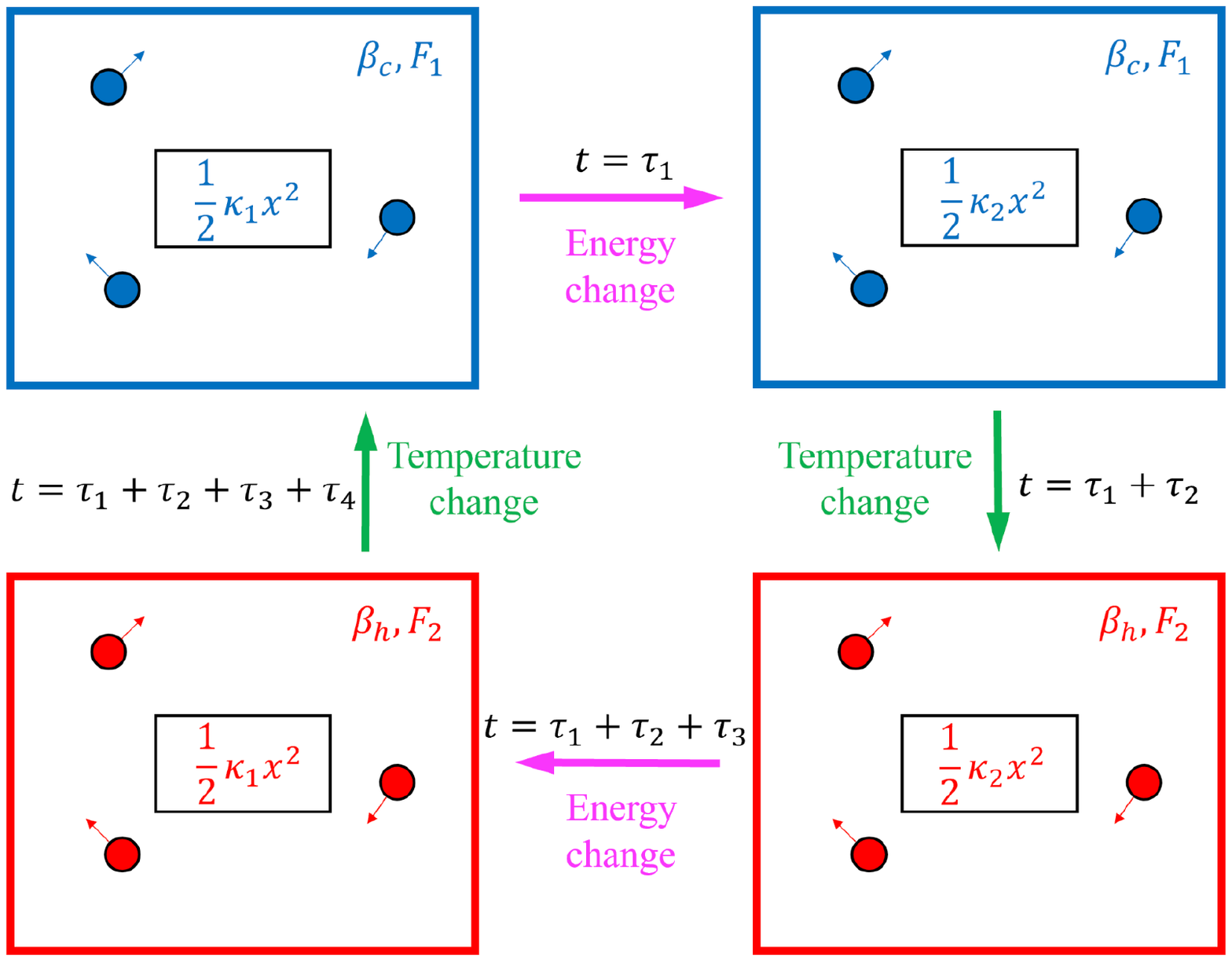}
\vspace{3mm}
\caption{Depiction of the protocol for a refrigerator with four steps. A colloidal particle subjected to a harmonic potential is represented by the equation $\kappa x^2/2$. The energy change is represented by a
change in the stiffness of the harmonic potential $\kappa$. The circles with arrows represent active particles and are a pictorial representation of the fact that the heat engine is active.}
\label{fig1}
\end{figure}

The time-periodic thermodynamic parameters follow the protocol depicted in Fig. \ref{fig1}. The duration of each step is denoted $\tau_k$ (k=1,2,3,4), and the total period is $\tau=\tau_1+\tau_2+\tau_3+\tau_4$.  The time-dependence of 
the thermodynamic parameters is as follows. The inverse temperature is
\begin{equation}
\beta(t)\equiv  
\begin{cases}
        \beta_c & \text{if } 0\le t <\tau_1+\tau_2 \\
        \beta_h & \text{if } \tau_1+\tau_2\le t <\tau.
 \end{cases}
\end{equation}
In this equation the subscript c (h) refers to cold (hot), which characterizes the inequality $\beta_h\le \beta_c$. The energy  reads
\begin{equation}
E_i(t)\equiv  
\begin{cases}
        E^{(1)}_i & \text{if } 0\le t <\tau_1 \\
        E^{(2)}_i & \text{if } \tau_1\le t <\tau_1+\tau_2+\tau_3\\
        E^{(1)}_i &  \text{if } \tau_1+\tau_2+\tau_3\le t <\tau. \\
 \end{cases}
 \label{eqprotE}
\end{equation}
We consider two specific models with an energy that depends on a continuous variable $x$ instead of the discrete variable $i$. The energy of these models 
is a harmonic potential $\kappa x^2/2$ and its dependence on the protocol represented 
by the superscripts in this equation appears as a subscript in the stiffness of the harmonic potential $\kappa$, as shown in Fig. \ref{fig1}. For the presentation of the second law from \cite{datt22} in this section, 
we keep the more general framework with discrete states. The thermodynamic affinity that makes the heat engine active is 
\begin{equation}
F(t)\equiv 
\begin{cases}
        F_1 & \text{if } 0\le t <\tau_1+\tau_2 \\
        F_2 & \text{if } \tau_1+\tau_2\le t <\tau.
 \end{cases}
\end{equation}
This form for $F(t)$ means that during the  part of the period when the temperature is cold, the affinity is $F_1$, and during the part of the period when the temperature is hot, the affinity is $F_2$.

The average dissipated heat per period in the cold reservoir is written as 
\begin{equation}
Q_c\equiv \int_{0}^{\tau_1+\tau_2} dt\sum_{i<j}  J_{ij}(t) [- \Delta E_{ij}(t)].
\label{eqqc} 
\end{equation} 
where $J_{ij}(t)$ is defined in Eq. \eqref{eqprobcurr}. The integration limits from $0$ to $\tau_1+\tau_2$ correspond to the time-interval when
the inverse temperature is $\beta_c$ and  $J_{ij}(t) [- \Delta E_{ij}(t)]$ is minus the rate of energy change due to a jump from $i$ 
to $j$. This ``passive'' heat does not take into account the heat dissipation directly associated with the affinity $F(t)$ that makes the heat engine active. 
Similarly, the average dissipated heat per period in the hot reservoir reads
\begin{equation}
Q_h\equiv \int_{\tau_1+\tau_2}^\tau dt\sum_{i<j}  J_{ij}(t)[- \Delta E_{ij}(t)].
\label{eqqh} 
\end{equation}

These heats can be written in a different from that is convenient for calculations. Using the relations $dP_i/dt= -\sum_jJ_{ij}(t)$, $dP_j/dt= \sum_iJ_{ij}(t)$, and the 
fact that there are only two piecewise changes in the protocol at times $\tau_1$ and $\tau_2$, we can write Eq. \eqref{eqqc} in the form 
\begin{equation}
Q_c=  \sum_iE_i^{(1)}[P_i(0)-P_i(\tau_1)]+\sum_iE_i^{(2)}[P_i(\tau_1)-P_i(\tau_1+\tau_2)] .
\label{eqqc2} 
\end{equation}
Similarly, Eq. \eqref{eqqh} can be written in the form 
\begin{equation}
Q_h= \sum_iE_i^{(2)}[P_i(\tau_2+\tau_1)-P_i(\tau_3+\tau_1+\tau_2)]+\sum_iE_i^{(1)}[P_i(\tau_3+\tau_2+\tau_1)-P_i(0)] ,
\label{eqqh2} 
\end{equation}
where we used $P_i(\tau)=P_i(0)$.

The average work per period  exerted on the refrigerator is given by
\begin{equation}
W\equiv \int_0^\tau dt\sum_iP_i(t)\frac{d}{dt}E_i(t).
\end{equation}
For the piece-wise protocol with the energy given by Eq. \eqref{eqprotE}, this expression becomes 
\begin{equation}
W\equiv \sum_i[P_i(\tau_1)-P_i(\tau_3+\tau_2+\tau_1)][E^{(2)}_i-E^{(1)}_i].
\label{eqW}
\end{equation}
These quantities fulfill a refined first law that does not involve the heat dissipation associated with the affinity $F(t)$ \cite{datt22}. This refined first law can be directly obtained from Eq. \eqref{eqqc2}, Eq. \eqref{eqqh2}, and  Eq. \eqref{eqW}.
It reads 
\begin{equation}
W=Q_c+Q_h.
\label{eqfirstlaw}
\end{equation} 
We reiterate that this refined first law implies that the work is only related to the contributions of heat dissipation that do not directly depend on the underlying dissipation associated with $F(t)$, in the case of work extraction corresponding to negative $W$, the source of this extracted work must come from these two heat contributions. We show next that an active system has an extra term in the second law that allows for unusual phases.

\subsection{Second law}

The probability distribution $P^S_i(t)$ plays an important role in the second law for cyclic active systems. This probability is the stationary distribution the system would reach if the protocol 
were frozen at time $t\in[0,\tau]$. In terms of the master equation \eqref{eqM},    $P^S_i(t)$ is the solution of 
\begin{equation}
\sum_j\left(P^S_j(t)w_{ji}(t)-P^S_i(t)w_{ij}(t)\right)=0.
\end{equation}
If $F(t)=0$, these transition rates, which depend on $F(t)$, are such that the solution of this equation is the equilibrium distribution $P^S_i(t)=P^{eq}_i(t)\propto \textrm{e}^{-\beta(t)E(t)}$. 

The second law for cyclic active systems contains an extra term, in comparison to the standard statement of the second law, which reads \cite{datt22} 
\begin{equation}
\hh\equiv  -\int_0^\tau dt \sum_iP_i(t)\frac{d}{dt}\left(\ln\frac{P^S_i(t)}{P^{eq}_i(t)}\right).
 \label{eqhh}
\end{equation}
If the heat engine is passive, i.e. $F(t)=0$ during the whole period, then $P^S_i(t)=P^{eq}_i(t)$ and this term is $\hh=0$.  

The appropriate second law for cyclic active systems is obtained from the so-called excess entropy \cite{seif12,hata01,trep04,spec05a,cher06,espo10b,espo10d,espo10f,pere11,saga11b}. 
The average excess entropy change  per period is given by
\begin{equation}
\Delta S_{\textrm{ex}}\equiv  \int_0^\tau dt \sum_{i<j} J_{ij}(t)\ln\frac{P^S_j(t)}{P^S_i(t)}= \beta_hQ_h+\beta_cQ_c+\hh\ge 0,
 \label{eqex}
\end{equation}
where the second equality can be derived from the generalized detailed balance relation Eq. \eqref{eqgdb}, as demonstrated in \cite{datt22}.  The extra term $\hh$, which becomes zero for a passive heat engine, 
in this expression for the excess entropy is the quantity that allows for the unusual phases of active systems. A key aspect of this second law is that it does not contain the energy dissipation directly associated with 
the affinity $F(t)$. Instead, this affinity makes the heat engine active and generates the entropic term $\hh$. This term is associated with the shift from the equilibrium Boltzmann distribution for $F(t)=0$ to the nonequilibrium steady 
state distribution  the system would reach if the protocol were frozen for $F(t)\neq0$. In other words, the amount of dissipation associated with $F(t)$  is not really important for an active system, the important aspect is 
how probability distributions are shifted, as quantified by $\hh$. This second law, together with an exact formula for $\hh$ in terms of a time derivative of Kullback-Leibler divergences, is discussed in further detail in \cite{datt22}.

\subsection{Coarse-graining and continuous variables}

An important issue for cyclic active systems is that of coarse-graining. The state of the system represented by the 
variable $i$ is divided into a visible variable $x$ and a hidden variable $a$, as represented by the equation $i=(x,a)$. For instance, $x$ 
can represent the position of a colloidal particle subjected to a harmonic potential. The hidden variable $a$ can represent the state 
of several active particles (such as the bacteria in the experiment from \cite{kris16}) that interact with the colloid labelled 
by $x$. The dynamics of $i=(x,a)$ is Markovian, however, the dynamics of the visible variable $x$ alone does not need to be Markovian. 
Under coarse-graining, the thermodynamic quantities can change, as they have to be calculated with the sole observation of $x$.
As demonstrated in \cite{datt22}, the structure of the second law in Eq. \eqref{eqex} remains the same, with the extra term $\hh$ in Eq. \eqref{eqhh} changed to 
\begin{equation}
\hh^{CG}\equiv  -\int_0^\tau dt \sum_xP_x(t)\frac{d}{dt}\left(\ln\frac{P^S_x(t)}{P^{eq}_x(t)}\right),
 \label{eqhhcg}
\end{equation}
where $P^S_x= \sum_aP^S_i$ and $P^{eq}_x= \sum_aP^{eq}_i$. For the models considered here, heat and work remain the same under coarse-graining, since they only depend on the variable $x$. 
The expression above  has a summation over $x$ instead of a summation over $i$ in Eq. \eqref{eqhh}. For the two models we consider 
here, we actually calculate the coarse-grained term in Eq. \eqref{eqhhcg}. In order to avoid carrying out the 
superscript, we denote the coarse-grained term in Eq. \eqref{eqhhcg} by $\hh$ for the rest of the manuscript.

The variable $x$ of the models we consider here is the  continuous position of a colloid, instead of the discrete variables considered in this section. 
However, these models can be described by overdamped Langevin dynamics. Hence, they can be obtained as a simple continuous limit 
of the more general discrete framework discussed in this section. In this continuum limit, the master equation \eqref{eqM} becomes the 
Fokker-Planck equation for overdamped dynamics. Expressions for the continuous variable become an integral over $x$ instead of 
a summation over $x$.

\section{Performance of an active refrigerator}
\label{sec3}

The presence of the term $\hh$ allows for new phases in cyclic active refrigerators, when compared to standard passive ones. These phases can be inferred by inspecting Eq.  \eqref{eqex}. 
Using the first law in Eq. \eqref{eqfirstlaw}, we rewrite the second law in Eq. \eqref{eqex} as 
\begin{equation}
\COPc(W+\beta_h^{-1}\hh)- (-Q_c)\ge 0,
\label{eqsecondlaw}
\end{equation}
where  
\begin{equation}
\COPc\equiv \beta_h/(\beta_c-\beta_h)
\label{eqCOPc}
\end{equation}
is the coefficient of performance of an ideal Carnot refrigerator and $-Q_c$ is the average heat per period taken from the cold reservoir.
 
The pseudo-coefficient of performance is defined as 
\begin{equation}
\COPps\equiv (-Q_c)/W. 
\label{eqCOPps}
\end{equation}
This quantity fulfills the bound $\COPps\le\COPc$ if $\hh=0$, i.e., if the refrigerator is passive. However, if the refrigerator is active, due to the presence of 
the $\hh$ term in the second law, the pseudo-coefficient of performance $\COPps$ can, in principle, be larger than the Carnot bound $\COPc$.  In fact, we will show 
with specific models that $\COPps$ can indeed be larger than $\COPc$.  A proper definition of a coefficient of performance 
must take the $\hh$ term into account and reads 
\begin{equation}
\COP\equiv (-Q_c)/(W+\beta_h^{-1}\hh)\le \COPc,
\label{eqCOP}
\end{equation}
where the inequality follows from Eq. \eqref{eqsecondlaw}. For a passive refrigerator with $\hh=0$, we have the equality $\COP=\COPps$.

Close inspection of this last equation shows that it is still possible to take heat from the cold reservoir with $W=0$. Hence, an active 
refrigerator can function like the original thought  experiment of Maxwell's demon, with heat flowing from cold to hot and no work input. The coefficient of 
performance of an active refrigerator given in Eq. \eqref{eqCOP} remains a well-defined quantity with the denominator given by $\beta_h^{-1}\hh$.

Beyond this Maxwell's demon phase, an active refrigerator can operate in a hybrid phase where it functions as both a heat engine and 
a refrigerator, corresponding to positive heat taken from the cold reservoir $-Q_c$ and positive extracted work $-W$. For this phase we rewrite the second 
law in Eq. \eqref{eqsecondlaw} in the following form,
\begin{equation}
\hh- \beta_h(-W)-(\beta_c-\beta_h)(-Q_c)\ge 0,
\label{eqsecondlaw2}
\end{equation}
In this hybrid phase, the coefficient of performance in Eq. \eqref{eqCOP} could even be negative and does not quantify the performance of the system. 
We propose the hybrid efficiency    
\begin{equation}
\xi\equiv \frac{\beta_h(-W)+(\beta_c-\beta_h)(-Q_c)}{\hh}\le 1
\label{eqpsi}
\end{equation}
as a possible quantifier of the performance of the system in the hybrid phase. Note that this quantity remains meaningful within the Maxwell's demon phase with $W=0$,
for which $\COP= \COPc\xi$. 

The phase corresponding to a heat engine is not analyzed in detail here since this phase has already been considered in \cite{datt22}. However, we do plot full phase diagrams
that include the region where the model is in the heat engine phase. We just specify that in the heat engine phase  $-Q_h$, $Q_c$,  and $-W$ are positive. In words, 
the system takes heat from the hot heat bath. Part of this heat becomes extracted work and part of it is released  in the cold reservoir.

In summary, from a simple analysis of the second law for cyclic active systems, we conclude that they can, in principle, display three different new phases. One is
a refrigerator with $\COPps>\COPc$, the second phase is Maxwell's demon, and the third phase is the hybrid (refrigerator and heat engine). We realize all these 
phases in two models, with an emphasis on the analysis of the role of the thermodynamic affinity or activity for the realization of these phases.

Let us explain the dimensions in our calculations below, where we set $\beta_h=1$. The hot temperature $T_h$ is an arbitrary temperature in Kelvin. The cold temperature  in Kelvin is $T_c= T_h\beta_c^{-1}$.
The numerical values of the energies per period $W$, $Q_c$, and $Q_h$ are all in units of $k_B T_h$, where $k_B$ is Boltzmann's constant.

\section{Results for the first model}
\label{sec4}

For our first model, we consider an active particle in one dimension subjected to a harmonic potential $V(x)=\kappa x^2/2$, which was used to realize an active heat engine in \cite{datt22}. 
Similar models have been studied in \cite{zaki17,saha19,holu20,kuma20}. This active particle is subjected to a force with a random direction that changes between 
$\pm F$ at a rate $\gamma$. The protocol for the model is illustrated in Fig. \ref{fig1}. For the calculations below, we perform the substitution $\kappa_1=\kappa$, 
$\kappa_2=\kappa+\Delta\kappa$, $F_1=F$, and $F_2=F+\Delta F$.

The exact solution of the model is explained in \ref{app1}. The exact expressions for heat and work are given by 
\begin{equation}
-Q_c=\frac{1}{2} \left(\frac{1}{\beta _c}-\frac{(\text{$\Delta $F} \kappa -\Delta \kappa  F) (\text{$\Delta $F} \kappa +\Delta \kappa  F+2 F \kappa )}{\kappa  (\Delta \kappa +\kappa )^2}-\frac{\kappa }{(\Delta \kappa +\kappa ) \beta
   _h}\right),
\label{eqqcexp}   
\end{equation}
where $-Q_c$ represents heat taken from the cold reservoir, and
\begin{equation}
W=\frac{1}{2} \Delta \kappa  \left(\frac{1}{\kappa  \beta _c}+\frac{F^2}{\kappa ^2}-\frac{\Delta \kappa +(\text{$\Delta $F}+F)^2 \beta _h+\kappa }{(\Delta \kappa +\kappa )^2 \beta _h}\right).
\label{eqwexp}
\end{equation}
The extra term $\hh$ does not have a closed expression, however, it can be written in terms of an integral 
shown in Eq. \eqref{eqhhmodel} in \ref{app1}. For specific values of the parameters, we can evaluate $\hh$ by performing this integral.

 Let us first consider the case of a passive refrigerator, which corresponds to $F=\Delta F=0$. In this case, the heat taken from the cold reservoir  in eq. \eqref{eqqcexp} becomes 
 $-Q_c= [\beta_c^{-1}-\beta_h^{-1}\kappa/(\kappa+\Delta\kappa)]/2$. Hence,  the passive system operates as a refrigerator, i.e.,  $-Q_c$ is positive, if the condition 
 \begin{equation}
 \kappa/(\kappa+\Delta\kappa)\le \beta_h/\beta_c
 \label{eqpassivecond}
 \end{equation}
 is fulfilled. For this passive case $\COP=\COPps$, since $\hh=0$, and $\COPps= \kappa/\Delta\kappa$, which can be readily obtained from the expressions above for heat and work. From the condition for 
 the system to function as a refrigerator that reads $\kappa/(\kappa+\Delta\kappa)\le \beta_h/\beta_c$, we see that the $\COP=\COPps\le \COPc$ for a passive refrigerator.
 
 \begin{figure}
\centering
\subfigure[]{\includegraphics[width=51mm]{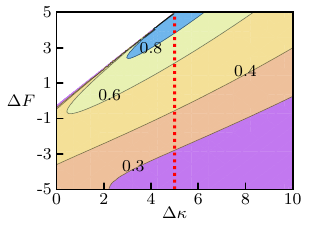}\label{fig2a}}
\subfigure[]{\includegraphics[width=51mm]{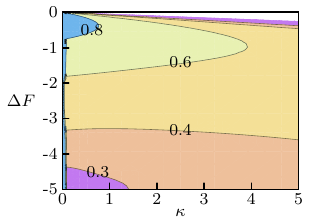}\label{fig2b}}
\subfigure[]{\includegraphics[width=51mm]{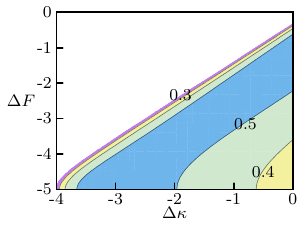}\label{fig2c}}
\vspace{-3mm}
\caption{Performance of an active refrigerator. The number in the lines separating two regions indicate the values of the $\COP$ or the parameter $\xi$ for the hybrid phase. (a) Refrigerator phase for $\kappa= 5$ and 
$F=5$, where $\COPps>\COPc$ for $\Delta \kappa$ less than the value 
indicated by the red dotted line. (b) Maxwell's demon phase with $\Delta \kappa= 0$ and $F=5$. (c) Hybrid phase for $\kappa=5$ and $F=5$. The inverse temperatures are set to $\beta_c=2$ and $\beta_h=1$ in all three figures. }
\label{fig2}
\end{figure}

 As shown in Fig. \ref{fig2a}, the  ratio $\COPps= -Q_c/W$ can exceed the bound $\COP_c$ if the refrigerator is active. The expression for the pseudo-coefficient of performance  $\COPps= (-Q_c)/W$ is 
 simply $\COPps=\kappa/\Delta\kappa$, which can be obtained from Eq. \eqref{eqqcexp} and Eq. \eqref{eqwexp}. The limiting condition for a positive $-Q_c$ in a passive refrigerator, given in  Eq. \eqref{eqpassivecond}, 
 does not hold for an active refrigerator, since $-Q_c$ also depends on the variables $F$ and $\Delta F$. Therefore, by setting $\Delta\kappa< \kappa/\COPc$ it is possible to have an active refirgerator
 with $\COPps>\COPc$, as indicated by the red dotted line in Fig. \ref{fig2a}.  Hence, we  have shown with exact calculations  that indeed an active refrigerator can have a pseudo-coefficient 
 of performance, which does not take the $\hh$ term into account, larger than the bound given by an ideal Carnot refrigerator. The $\COP$ that takes the $\hh$ term into account is bounded by $\COPc$, as shown in Fig. \ref{fig2a}.

The Maxwell’s demon phase in this model is obtained by setting $\Delta \kappa=0$, which leads to $W=0$, as shown in Eq. \eqref{eqwexp}. The expression for the heat taken from the cold reservoir in Eq. \eqref{eqqcexp}  becomes
\begin{equation}
-Q_c=\frac{1}{2} \left(-\frac{\Delta F(2F+\Delta F)}{\kappa}-(\beta_h^{-1}-\beta_c^{-1})\right),
\end{equation}
This expression has a clear physical interpretation. The system operates as Maxwell's demon with heat extraction from the cold reservoir if the active contribution is larger than the passive term, i.e.,
 \begin{equation}
-\frac{\Delta F(2F+\Delta F)}{\kappa}\ge \beta_h^{-1}-\beta_c^{-1}.
\end{equation}
Hence, $\Delta F<0$ is a necessary condition for the system to operate as Maxwell's demon. From the equation, a negative $F$ with a positive $\Delta F$ could also fulfill the condition. However,
since the model is symmetric with respect to $F$ (the active particle changes between $\pm F$ at rate $\gamma$), a positive $\Delta F$ with a negative $F$ is analogous to a positive $F$ with a negative $\Delta F$.
In conclusion, the affinity during the hot part of the cycle $F_2=F+\Delta F$ has to be smaller than the affinity during the cold part of the cycle $F_1=F$, for the realization of the Maxwell's demon phase within this model. 
In Fig. \ref{fig2b}, we show the $\COP$ that is bounded by $\COPc$ for this Maxwell's demon phase.

Finally, if we now set $\Delta \kappa$ to be negative, it is possible to have both the extracted work $-W$ and the heat taken from the cold reservoir $-Q_c$  positive, which corresponds to the hybrid phase. This result 
is illustrated in Fig. \ref{fig2c}, where we plot the hybrid efficiency  $\xi$ from Eq. \eqref{eqpsi}. It is the active contributions, due to $F$ and $\Delta F$, in the analytical expressions in Eq. \eqref{eqqcexp} and Eq. \eqref{eqwexp}
that allow for this hybrid  phase.

\begin{figure}
\centering
\includegraphics[width=85mm]{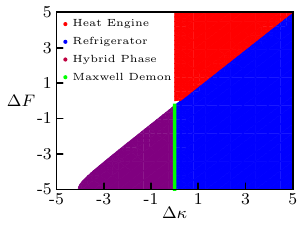}
\vspace{-3mm}
\caption{Phase Diagram for the exactly solvable model. The parameters are set to $\kappa= 5$, $F=5$, $\beta_c= 2$, and $\beta_h=1$.}
\label{fig3}
\end{figure}

The full phase diagram of this model is shown in Fig. \ref{fig3}. Our results demonstrate the realization of the phases speculated in Sec. \ref{sec3} in an exactly solvable model. We now show that these phases can also be 
realized in a more realistic model that is closer to the original experiment for an active heat engine from \cite{kris16}.

\section{Results for the second model}
\label{sec5}

The second model is a passive particle subjected to a harmonic potential that interacts with several active particles, similar to the situation in the experiment from \cite{kris16}. 
 In our notation, the variable $x$ labels the position of the passive particle and the variable $a$ the state of the 
active particles. Another difference with the previous model is that this model is two-dimensional.  This model was used in \cite{datt22} to realize a cyclic active heat engine and here 
we use it to realize a refrigerator.

The full definition of the model, together with the numerical method used to analyze it, is given in \ref{app2}. The variables in the model that are important to our analysis in the main text are 
the stiffness of the harmonic potential $\kappa$, which is the same as in the previous model, and the magnitude of the active velocity of the active particles $u$. This parameter $u$ plays a role similar 
to $F$ in the previous model. If $u=0$, the system would reach an equilibrium stationary state and the heat engine is passive. The active particles have speed $u$ and the direction of the velocity 
is a random angle, similar to the affinity $F$ in the first model that changes between positive and negative directions randomly. This parameter $u$ is set as $u_1$ for the part of the period when the temperature 
is cold and $u_2$ for the part of the period when the temperature is hot. The inverse temperatures are set to $\beta_c=1.2$ and $\beta_h=1$.

\begin{figure}
\centering
\subfigure[]{\includegraphics[width=51mm]{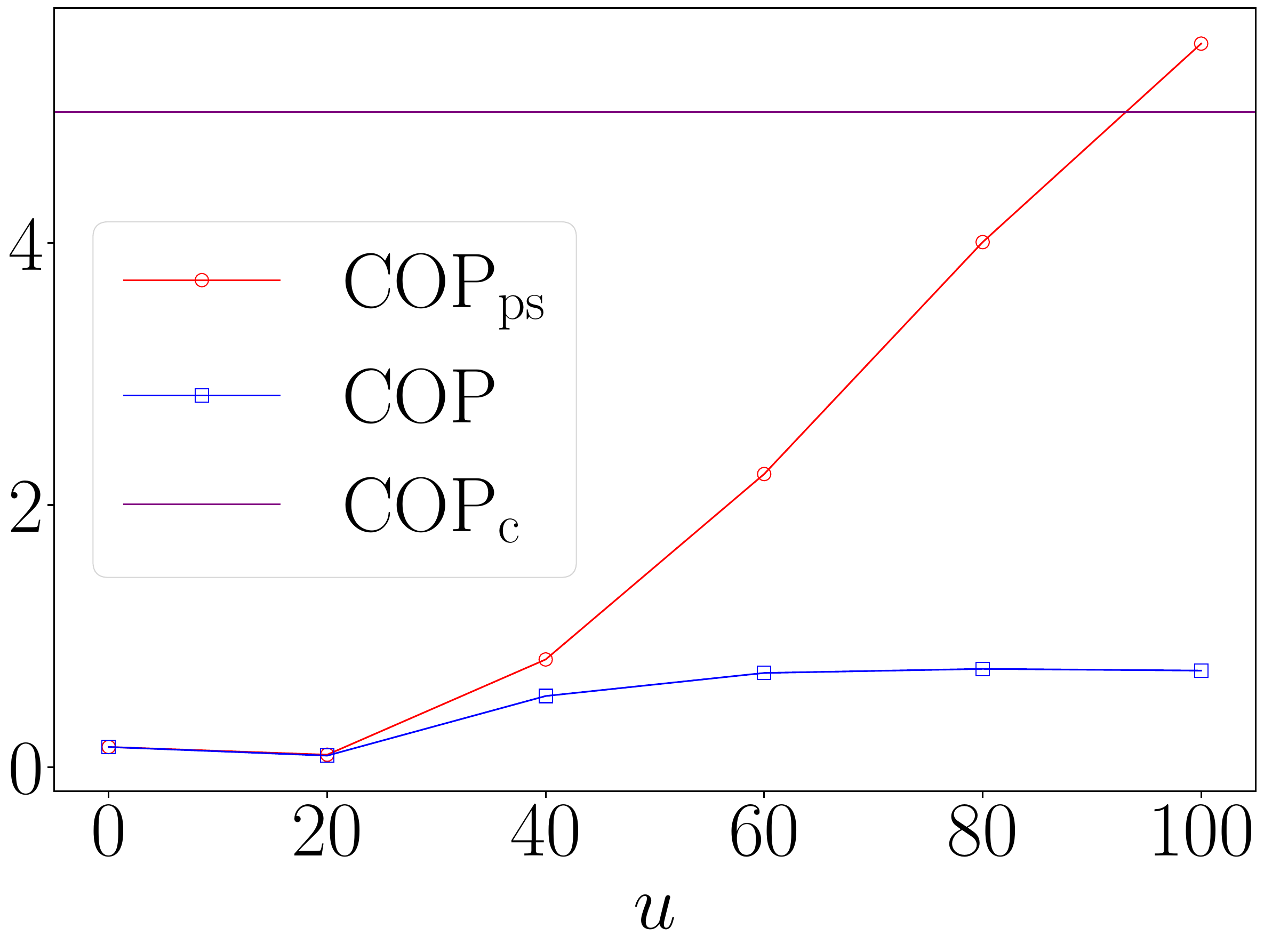}\label{fig4a}}
\subfigure[]{\includegraphics[width=51mm]{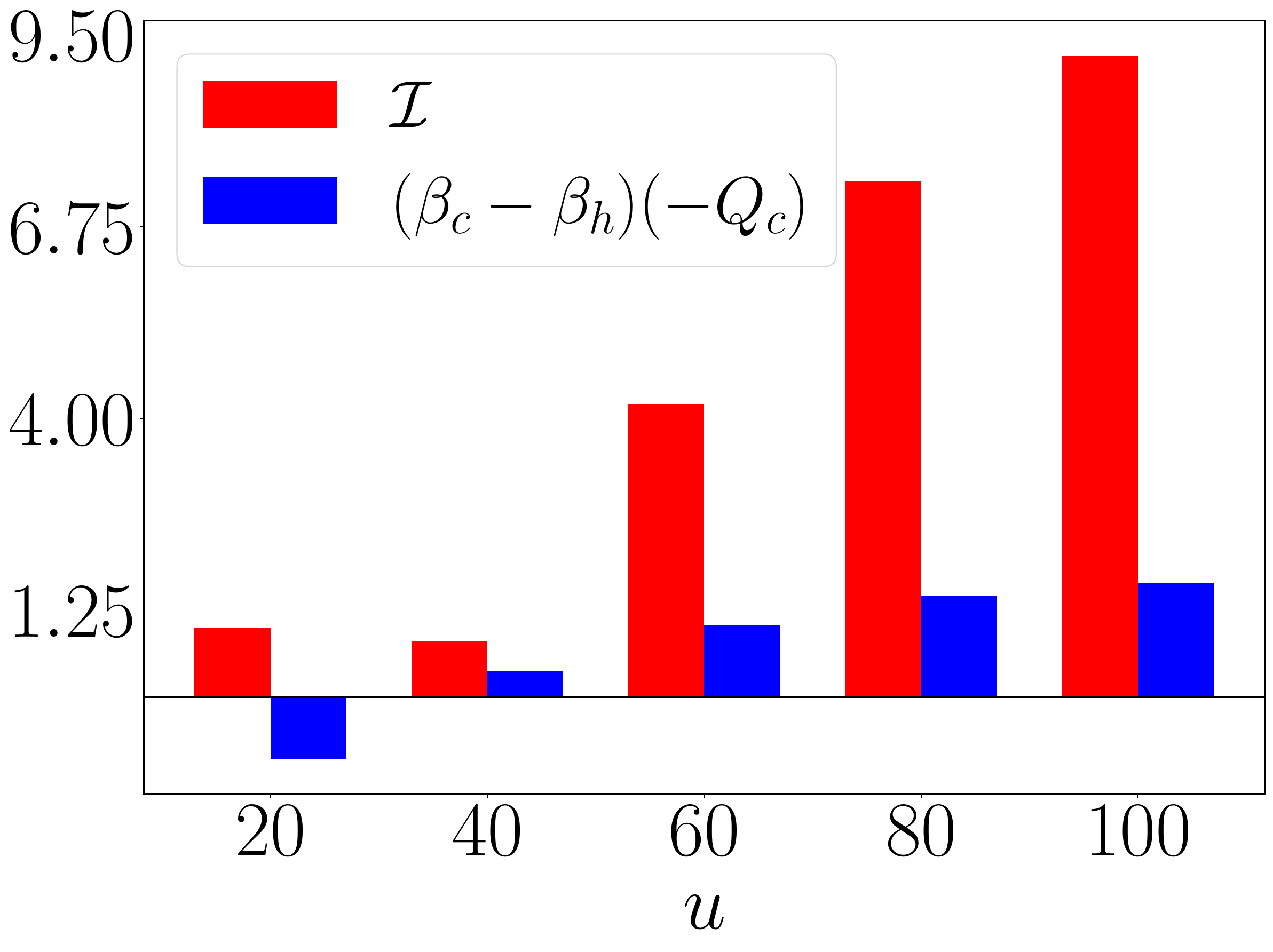}\label{fig4b}}
\subfigure[]{\includegraphics[width=51mm]{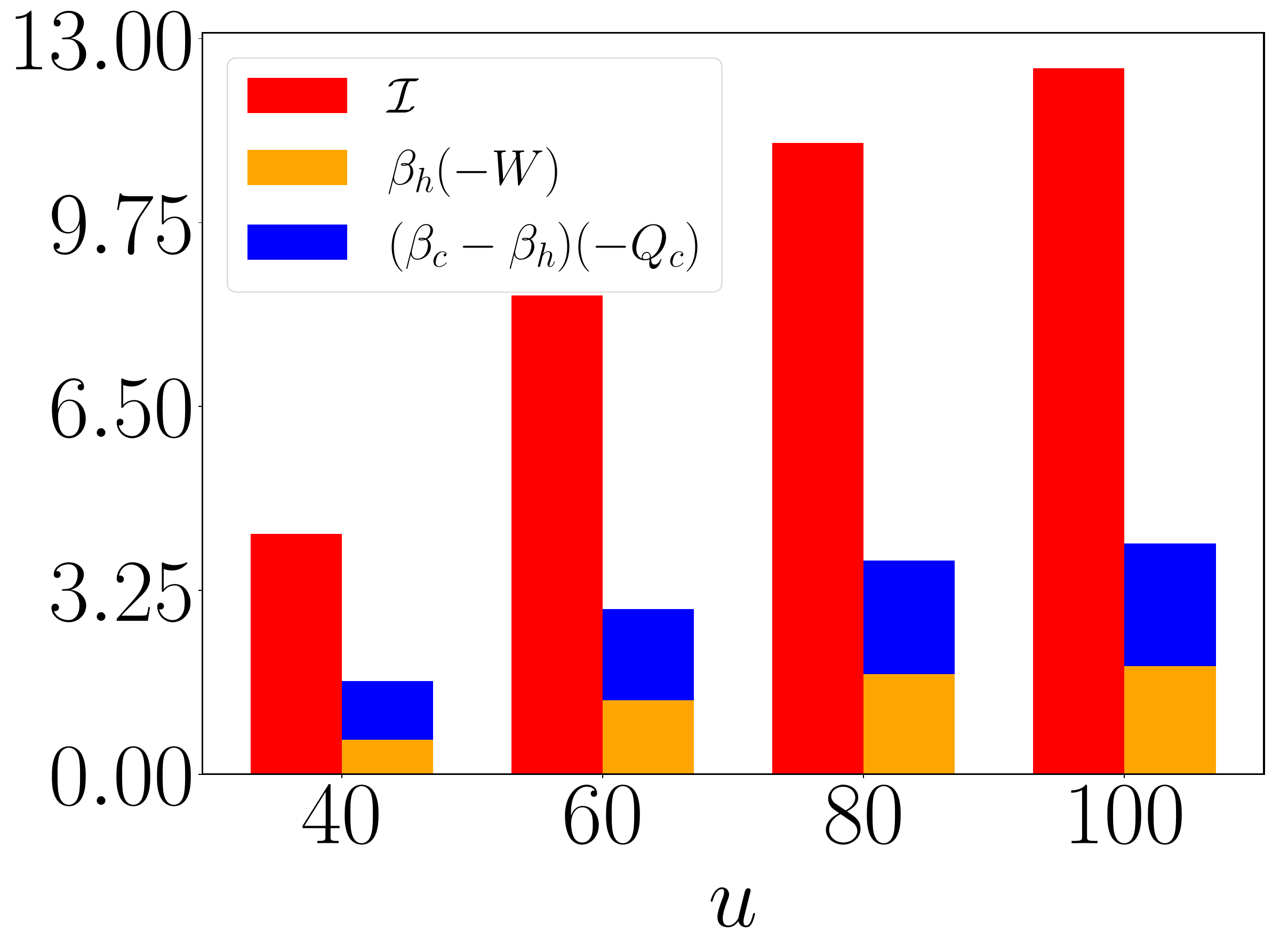}\label{fig4c}}
\vspace{-3mm}
\caption{Performance of of an active refrigerator. (a) Refrigerator phase with $\COPps>\COPc$. The parameters are set to $\kappa_1= 416.5$, $\kappa_2=83.3$,  $u_1= u$, and $u_2= 15 u$.  (b) Realization of Maxwell's demon with positive $-Q_c$ and no work input. 
The parameters are set to $\kappa_1 = \kappa_2 = 83.3$,  $u_1 = u$, and $u_2=10u$.(c) Realization of hybrid phase. The parameters are set to $\kappa_1= 141.6$, $\kappa_2=83.3$, $u_1=u$, and $u_2 = 15u$.}
\label{fig4}
\end{figure}

In Fig. 4, we show the realization of the three phases observed with the previous model. We see that $\COPps$ can be larger than $\COPc$ in Fig. \ref{fig4a}, the realization of the Maxwell's demon phase for $\Delta \kappa$ in Fig. \ref{fig4b},
and the hybrid phase in Fig. $\ref{fig4c}$. For this model, the condition that the activity is smaller during the hot part of the cycle, which implies $\Delta u= u_2-u_1<0$, is not a necessary condition for the onset of the two last phases, since they are realized with 
$\Delta u>0$ in Fig. \ref{fig4}. This result is in contrast to the previous model that required a negative $\Delta F$ for the onset of these phases. 

\begin{figure}
\centering
\includegraphics[width=85mm]{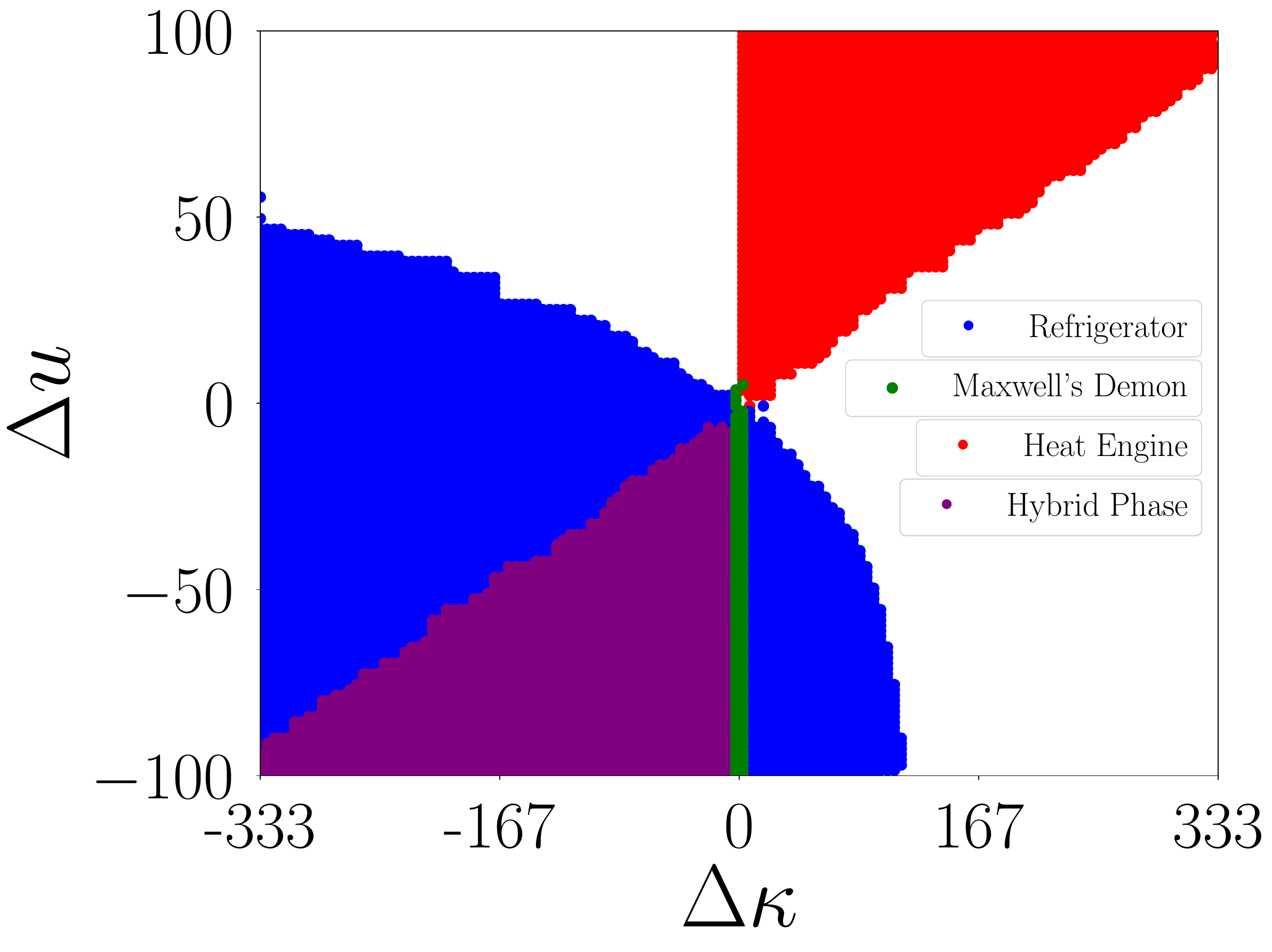}
\vspace{-3mm}
\caption{Phase Diagram for the second model in the plane $(\Delta\kappa= \kappa_2-\kappa_1,\Delta u=u_2-u_1)$. Parameters are set in the following way: we fixed the sums  $\kappa_1+\kappa_2= 333.2$ and $u_1+u_2=100$. }
\label{fig5}
\end{figure}

The full phase diagram of the model is shown in Fig. \ref{fig5}. In Summary, we have shown that all three novel phases of a cyclic active refrigerator can be 
realized in a more realistic setting that is quite close to the experiment reported in \cite{kris16}.

\section{Conclusion}
\label{sec6}

We introduced the concept of  cyclic active refrigerators. They display three new phases in relation to passive systems. First, there is a phase where the ratio of heat taken from the cold reservoir divided by the work input can be larger 
than the coefficient of performance of an ideal Carnot refrigerator. Second, there is a Maxwell's  demon phase where the refrigerator can function without any work input. Third, there is a hybrid phase with the cyclic active 
system performing both tasks,  heat removal from the cold reservoir such as a refrigerator and  work extraction such as a heat engine. All these phases are possible due to the emergence 
of the $\hh$ term in the second law for cyclic active systems. This term is nonzero only if the refrigerator is active, i.e., if there is a nonzero affinity that would drive the system to a nonequilibrium stationary state if the protocol were frozen. 

Our results were obtained with two different models. The first exactly solvable allowed for a straightforward realization of these phases, with explicit expressions that allow us to observe the role of the thermodynamic affinity in 
an active system quite directly. The second model of a passive particle interacting with several active particles is a close setup to the experiment from \cite{kris16}. Our realization of these three phases with this more realistic model 
shows that all these phases of a cyclic active refrigerator can  be observed in experiments of a colloidal particle in a bath full of bacteria. Furthermore, we show with this more complex model that reduced activity at higher 
temperatures is not a necessary condition for the onset of the Maxwell’s demon phase and the hybrid phase, which is a restriction for the first model that requires a negative $\Delta F$.

From the point of view of information and thermodynamics, our work is also relevant. The original thought experiment of Maxwell's demon requires an external agent to perform measurement and feedback. We demonstrated  with a different 
approach that such a phase of extraction of heat from a cold reservoir without work input can happen in a more inclusive setup, without the need for an external agent. The underlying thermodynamic affinity that leads 
to  hidden dissipation that does not show up in the second law generates the $\hh$ term that allows for the realization of this phase. Mawell's demon is a particular case of more general  feedback-driven systems \cite{saga12}.
There are other approaches in information and thermodynamics that are more inclusive than feedback-driven systems. For instance,  a generic system interacting with a tape \cite{mand12,mand13,deff13} 
and the so-called bipartite systems \cite{hart14,bara14a,horo14}. Our original contribution here is that Maxwell's demon can be realized with yet another more inclusive approach.

Concerning future work, the investigation of cyclic active systems remains an interesting area of research. One possible direction is to investigate cyclic active refrigerators made of several degrees of freedom and the realization of these 
three phases in a macroscopic system.

{\noindent \textbf{Acknowledgements}}\newline We thank Patrick Pietzonka for helpful discussions. We  thank the financial support from the NSF through the grant  DMR-2424140.

{\noindent \textbf{Data availability statement}}\newline  The data that support
the findings of this study are available upon reasonable request from the authors.

\appendix

\section{Calculations for the first model}
\label{app1}

The full state of the system is determined by the continuous variable $x$ that gives the position of the particle and the discrete variable $a=\pm 1$. The
particle is active due to a force that has a direction that depends on the variable $a$, with a positive force $+F$ for $a=1$ and a negative force $-F$ for $a=-1$. The force changes between positive and negative with a rate $\gamma$ that is independent of position $x$. 

We consider a limit that allows for a straightforward calculation of the stationary distribution, which corresponds to a certain time-scale separation. The rate $\gamma$ for changing the direction of the force
 is much smaller than the rate for transitions that change the position $x$.
 Hence, for a given force, the system first reaches a stationary distribution associated with a fixed force and then changes to a different value of the force. Formally, the conditional distribution of $x$ given $a=\pm 1$ is then the stationary  distribution for a fixed force, i.e.,
\begin{equation}
P^S_{x|a}=\frac{\sqrt{\beta\kappa}}{\sqrt{2\pi}}\exp(-\beta F^2/(2\kappa)) \exp(-\beta (\kappa x^2/2-aFx)).
\end{equation}
For the variable $a$ alone, we just have a process changing between two values with the same rate $\gamma$. Therefore, the marginal distribution  of the variable 
$a$ is simply  $P^S_a=1/2$ for both values $a=\pm1$. We can now obtain the marginal distribution for $x$ using the joint distribution $P^S_{x,a}=P^S_{x|a}P^S_a$
and the relation $P^S_x= \sum_aP^S_{x,a}$, which is 
\begin{equation}
P^S_x\equiv P_x^{(\beta,\kappa,F)}= \frac{\sqrt{\beta\kappa}}{\sqrt{2\pi}}\exp(-\beta F^2/(2\kappa)) \exp(-\beta \kappa x^2/2)\cosh(\beta Fx),
\label{eqPSFirst}
\end{equation}
where we have introduced the notation $P_x^{(\beta,\kappa,F)}$ to keep the dependence of the three variables in the superscript.   

In possession of the stationary distribution in Eq. \eqref{eqPSFirst} we can now calculate the thermodynamic quantities $Q_c$, $Q_h$,  $W$ and $\hh$ if we make further assumptions about 
the time-scales of the protocol illustrated  in Fig. \ref{fig1}. We assume that $\tau_2\to 0$ and $\tau_4\to 0$, which implies that the temperature changes are instantaneous, i.e., much faster compared to
any transition rate of the system. The other times $\tau_1=\tau_3=\tau/2$ are assumed to be very long, such that the system has time to relax to a stationary state within this time. Hence, the periodic 
long-time limit probability fulfills $P_x(\tau/2)= P_x^{(\beta_c,\kappa_1,F_1)}$ and $P_x(\tau)= P_x^{(\beta_h,\kappa_2,F_2)}$. 

For the conditions explained in the paragraph above, we obtain the following expressions for the thermodynamic quantities. 
The average dissipated heat per period in the cold reservoir in Eq. \eqref{eqqc2} becomes 
\begin{equation}
Q_c=  \int_{-\infty}^\infty dx(\kappa_1x^2/2)(P_x^{(\beta_h,\kappa_2,F_2)}-P_x^{(\beta_c,\kappa_1,F_1)}) .
\label{eqqcfirstmodel} 
\end{equation}
In order to obtain this expression, we performed the following substitutions, $\sum_i\to \int dx$, $E_i^{(1)}\to \kappa_1 x^2/2$, $\tau_1= \tau/2$, $\tau_2=0$, 
$P_i(0)\to P_x^{(\beta_h,\kappa_2,F_2)}$, and $P_i(\tau_1+\tau_2)\to P_x^{(\beta_c,\kappa_1,F_1)}$. Similarly, from Eq. \eqref{eqqh2}, we otain
\begin{equation}
Q_h=  \int_{-\infty}^\infty dx(\kappa_2x^2/2)(P_x^{(\beta_c,\kappa_1,F_1)}-P_x^{(\beta_h,\kappa_2,F_2)}) .
\label{eqqhfirstmodel} 
\end{equation}
The expression for work in \eqref{eqW}
\begin{equation}
W\equiv \int_{-\infty}^\infty dx(P_x^{(\beta_c,\kappa_1,F_1)}-P_x^{(\beta_h,\kappa_2,F_2)})(\kappa_2-\kappa_1)x^2/2.
\label{eqW2}
\end{equation}
The explicit expressions in the main text,  given in Eq. \eqref{eqqcexp} and in Eq. \eqref{eqwexp}, are obtained by integrating these equations.
Finally, the expression for the $\hh$ term in Eq. \eqref{eqhhcg} for this model is 
\begin{equation}
\hh= \int_{-\infty}^{\infty} (P_x^{(\beta_h,\kappa_2,F_2)}-P_x^{(\beta_c,\kappa_1,F_1)})\{\ln[\cosh(\beta_hF_2x)]-\ln[\cosh(\beta_cF_1x)]\}.
\label{eqhhmodel}
\end{equation}

\section{Numerical simulations of the second model}
\label{app2}

The model of a passive particle interacting with active particles is defined as follows. The position of the passive particle in two dimensions is labeled by $\vec{x}=(x_1,x_2)$. This particle is 
subjected to the harmonic potential $V(x)=\kappa (x_1^2+x_2^2)/2$. The passive particle interacts with several active particles  via a quadratic interaction potential. 
The passive particle has radius $R$, while the active particles are assumed to be points with no size.
The position of active particle $m$ is denoted $\vec{a}_m$, and the interaction potential reads  
\begin{equation}
E_{\textrm{int}}(|\vec{a}_m-\vec{x}|)=
\left\{
	\begin{array}{ll}
		0  & \textrm{if }  |\vec{a}_m-\vec{x}|>R \\
		\kappa_\mathrm{rep}(R-|\vec{a}_m-\vec{x}|)^2/2 & \textrm{if } |\vec{a}_m-\vec{x}|<R
	\end{array}
\right.,
\label{eqrepu}
\end{equation}
where $\kappa_\mathrm{rep}>0$ for repulsive interaction.

The Langevin equation for the passive particle reads 
\begin{equation}
  \dot{\vec{x}}=-\mu_p \nabla_{\vec{x}}V(x)-\mu_p\sum_m\nabla_{\vec{x}}E_{\textrm{int}}(|\vec{a}_m-\vec{x}|)+\vec\xi(t)
  \label{eq:langevinp}
\end{equation}
 where $\mu_p$ is the mobility of the passive particle and $\vec\xi(t)$ corresponds to a Gaussian white noise with  intensity $\mu_p/\beta$. 

The active particle has active velocity with magnitude $u$  and direction $\vec{n}(\phi_m)$, which is a unit vector 
pointing in the direction of angle $\phi_m$. The random variable $\phi_m$ is characterized by Brownian diffusion with diffusion coefficient
$D_\mathrm{r}$ \cite{fily12,roma12}.   The Langevin equation for the active particles reads
\begin{equation}
\dot{\vec{a}}_m=u\vec{n}(\phi_m)-\mu_a\nabla_{\vec{a}_m}E_{\textrm{int}}(|\vec{a}_m-\vec{x}|)+\vec\zeta_m(t),
\label{eq:langevina}
\end{equation}
where $\mu_a$ is the mobility of the active particles and $\vec{\zeta}_m(t)$ is a Gaussian white noise with intensity
$\mu_a/\beta$. 

The protocol has the same four parts as the generic one shown in Fig. \ref{fig1}, with all times set to $\tau_1=\tau_2=\tau_3=\tau_4$. 
The speed $u$ of the active particles plays a role similar to $F$ for the first model,  this velocity is $u_1$ during the first half of the 
period and $u_2$ during the second half of the period.

The boundary conditions are as follows. The passive particle is unconstrained since it is unlikely to leave the region close to the minimum of 
the harmonic potential. The active particles are in a box with periodic boundary conditions. The dimensions of the box are  $-1\le x_1\le 1$ 
and  $-1\le x_2\le 1$. The radius of the passive particle is $R=0.2$.  Furthermore, the parameters of the model are set to the following values. 
The period is $\tau=20$, the total number of active particles is  30 ($m=1,2,\ldots,30$), the inverse temperatures are $\beta_c=1.2$ and $\beta_h=1$, 
the mobility of the passive particle is  $\mu_p=0.006$, the mobility of the active particle is $\mu_a=0.012$, and the diffusion coefficient for the changes in the direction of the active speed $u$ is given by $D_\mathrm{r}=1$. 
The strength of the repulsive potential in Eq. \eqref{eqrepu} 
is $\kappa_\mathrm{rep}=49980$ for the results shown in Fig \ref{fig4} and $\kappa_\mathrm{rep}=83300$ for the results shown in Fig \ref{fig5}. 
The parameters $\kappa_1$, $\kappa_2$, $u_1$, and $u_2$ are varied to obtain different results.

For the numerical simulations, we use the stochastic Euler method with the  Langevin equations above with a discrete time step that is small enough (given by $10^{-4}$). 
We calculate $Q_c$ and  $W$ from the stochastic trajectories generated in the numerical simulation. 
For the heat $Q_c$, whenever there is a change  in the position of the passive particle from $\vec{x}$ to $\vec{x}_{\textrm{new}}$, during the part of the period when the temperature is the cold one,
we sum up minus the change in potential energy $\kappa_1(x^2+y^2-x_{\textrm{new}}^2-y_{\textrm{new}}^2)/2$. We obtain the average heat in a period by dividing the total value from the long stochastic trajectory 
by the total number of periods in the simulation. For the work $W$, for every period, at time $\tau/4$ we sum up $(\kappa_2-\kappa_1)(x^2+y^2)/2$ 
and at time $3\tau/4$ we sum up $(\kappa_1-\kappa_2)(x^2+y^2)/2$, where $\vec{x}$ represents the position of the particle during a stochastic simulation at these times. We also divide the total value by the number
of periods to get the average work per period.

For the term $\hh$, we need the nonequilibrium stationary distribution, which is not known. There are four different stationary distributions corresponding to the four different parts of the period. We sample these four distributions 
numerically, by running four different simulations of the model with a fixed protocol and the corresponding parameters. The distributions can then be obtained by sampling a histogram in these simulations. This method is  explained
in more detail in \cite{datt22}. In possession of these distributions, we can calculate  the $\hh$ from a stochastic simulation of the model with a cyclic protocol. Whenever there is a piecewise change in the protocol during the stochastic trajectory,
 we increment the $\hh$ term in accordance with equation \eqref{eqhhcg}, where the $P^S(x)$ is obtained from the aforementioned numerical simulations  and $P^{eq}(x)$ is just a Gaussian associated with the harmonic potential.


\section*{References}



\end{document}